\documentstyle[aps]{revtex}
\tightenlines
\begin{document}
%\draft
\preprint{ }
\title{
Hofstadter-type energy spectra in lateral superlattices defined by periodic
magnetic and electrostatic fields
}
\author{Rolf R.\ Gerhardts and Daniela Pfannkuche}
\address{
Max-Planck-Institut f\"ur Festk\"orperforschung, Heisenbergstrasse 1,
D-70569 Stuttgart, Federal Republic of Germany
}
\author{Vidar Gudmundsson}
\address{ Science Institute,
University of Iceland, Dunhaga 3, IS-107 Reykjavik, Iceland 
}
\date{ 4 August 1995}
\maketitle
\begin{abstract}
We calculate the energy spectrum of an electron moving in a two-dimensional
lattice which is defined by an electric potential and 
an applied perpendicular magnetic field modulated by a periodic surface
magnetization. The spatial direction of this magnetization introduces complex
phases into the Fourier coefficients of the magnetic field. We investigate the
effect of the relative phases between electric and magnetic modulation on
band width and internal structure of the Landau levels.    
\end{abstract}
\pacs{PACS numbers: 73.20.-r, 73.20.Dx}
%
% \narrowtext
%
%
%\section{Introduction}
%
Energy spectra of electrons moving in the presence of a perpendicular
homogeneous magnetic field $B$ through a two-dimensional (2D) lattice are
determined by 
fascinating commensurability effects. For  a 2D
lattice described by the {\em electrostatic potential }
\begin{equation}
V(x,y)=V_x \cos (K x)+ V_y \cos (K y) \label{v2cos}
\end{equation}
 with lattice constant $a =2 \pi / K$, the
single-particle energy spectrum can be calculated 
 if the
magnetic flux $\Phi = B a^2$ through a unit cell of the
lattice is a rational multiple of the flux quantum $\Phi_0 = c h /e$,
\begin{equation}
\Phi / \Phi_0 \equiv 2 \pi /K^2 l^2 = p/q \, .
\label{fluxrat}
\end{equation}
Here $l=(c\hbar / eB)^{1/2}$ is the magnetic length, and $q$ and  $p$
are relative prime integers. 
If the amplitude of $V(x,y)$ is much smaller than the distance between
Landau levels, i.e. the cyclotron energy $\hbar \omega_c \equiv \hbar eB/mc$,
mixing of Landau levels can be neglected.
Then the resulting energy
spectrum consists of broadened Landau bands, each with the same internal
structure which, when
plotted versus the inverse flux ratio $\Phi_0 / \Phi$,   is
known as Hofstadter's butterfly\cite{Hofstadter76:2239},  see Fig.1(a). The
width of these Landau 
bands oscillates as a function of the Landau quantum number $n$ and the flux
ratio $\Phi / \Phi_0$. 

So far the gap structure predicted by Hofstadter's butterfly has not been
observed directly in experiments, but there is strong evidence that it plays an
important role for the magnetoresistance of the laterally modulated 2D
electron  systems in GaAs-AlGaAs heterostructures (for a review see
Ref.\cite{Pfannkuche92:12606}). Magnetoresistance measurements have been
performed on 1D and 2D lateral superlattices, with characteristically
different results. In
1D superlattice  so called ``Weiss
oscillations'' are observed, which can be 
understood from the oscillating width of the Landau bands. The 
dispersion of these bands  leads to a group velocity, the quantum 
analog of the classical  
guiding center drift of cyclotron orbits in the oscillating electric field.
As a consequence, an oscillatory contribution to the conductivity results,
which is called ``band conductivity'' and which is absent in homogeneous 2D
electron systems. In a 1D superlatttice the Weiss oscillations of the band
conductivity can also be calculated
quasi-classically\cite{Beenakker89:2020,Gerhardts92:3449} from the guiding
center drift. In weak 2D superlattices this band conductivity is
drastically suppressed, and this is attributed to the splitting of the Landau
bands into narrow subbands separated by large gaps, as predicted by the
Hofstadter spectrum. In such a spectrum the group velocities and, as a
consequence, the band conductivity are much smaller than in the corresponding
1D case.

The analog of the Weiss oscillations in 2D electron systems with
a {\em periodic magnetic field}, which had been
predicted\cite{Vasilopoulos90:393} and 
investigated in some detail\cite{Xue92:5986,Peeters93:1466} several years ago,
has  been experimentally observed\cite{Izawa95:706,Carmona95:3009,Ye95:3013} 
only recently. 
The magnetic modulation was achieved
by metallic strips on the surface, consisting of
superconducting\cite{Carmona95:3009} or
ferromagnetic\cite{Izawa95:706,Ye95:3013} material and exerting at low
temperatures stress on the underlying semiconductor, so that  the
magnetic modulation was found to be accompanied by a distinct electric
modulation. Systematic investigations\cite{Davies94:4800,Ye95:715} have shown
that this stress-induced electrostatic modulation always has a considerable
content of higher harmonics. Nevertheless, the experimental results for
unidirectional magneto-electric modulation are in reasonable agreement with
the theoretical calculations\cite{Xue92:5986,Peeters93:1466} which, in the
regime of the Weiss oscillations may even be replaced by their quasi-classical
limit\cite{Ye95:3013,Beenakker89:2020,Gerhardts:unpub}. 

Quite recently, magnetoresistance in 2D magneto-electric superlattices has
been investigated\cite{Ye95:subm,Ye:privcomm}. The lattice was defined by a
quadratic array of ferromagnetic posts deposited on the surface of a
GaAs-AlGaAs heterostructure, so that a combined magnetic and 
electric modulation results at low temperatures.  The observed oscillations
show distinct features which can not be explained by the quasi-classical
theory, which is sufficient to explain the corresponding results for 1D
modulations. Therefore the question arises if, similar to the case with purely
electric modulation, a Hofstadter-type energy spectrum suppresses the
modulation-induced band conductivity and how Hofstadter's butterfly is
modified. We will investigate this question for the most
simple modulation model which includes the experimentally relevant
symmetries. 

In the experiment, the ferromagnetic posts can be magnetized parallel or
perpendicular to the plane of the 2D electron gas. As a simple model for  the
magnetic field resulting for an arbitrary direction of the periodic
magnetization, we 
consider the field of identical magnetic dipoles $\vec{m}$ located at the
lattice sites 
$(n_x a, n_y a,0)$ in the plane $z=0$. It is a straightforward exercise to
calculate the Fourier expansion of the  resulting periodic
magnetic field in a plane $z <0$, where the 2D electron gas is assumed to be
located. For the $z$ component one obtains 
\begin{equation}
B_z(\vec{r};z)= \sum _{\vec{q}} B_{\vec{q}}(z) \, e^{i \vec{q}
  \cdot \vec{r}}\, , \label{fourier}
\end{equation}
where
\begin{equation}
B_{\vec{q}}(z) =\frac{2\pi}{a^2}   \left(i \, \vec{m} \cdot \vec{q}
+ m_z q \right)\, e^{ -q |z|} \, , 
\label{dipolfeld}
\end{equation}
with $q=| \vec{q} \,|$,  $\vec{q} =  (n_x K,n_y K,0)$ and integers $n_x$ and
$n_y$. Thus the complex phase of the Fourier coefficients $B_{\vec{q}}(z)$ is
determined by the orientation of the magnetic moments on the surface, a result
which remains 
valid beyond this simple dipole approximation. Also the exponential damping of
higher Fourier components is typical for magneto- and electrostatic fields
satisfying Laplace's equation.

We now assume a strictly 2D model
for the electron gas and neglect the in-plane components of the periodic
magnetic field together with the $z$ dependence of $B_z(\vec{r})$. To describe
$B_z(\vec{r})$ in the plane of the electrons, we take Eq.~(\ref{dipolfeld})
for the fixed $z$ value at this plane and  use in the 2D Hamiltonian
the vector potential $\vec{A} (\vec r)$ with Fourier coefficients
$\vec A _{\vec q} =i (q_y, -q_x,0) B_{\vec q} / q^2$, which commutes with the
momentum operator. For the homogeneous external magnetic field $B_0$ we use
the 
Landau gauge $\vec{A}^{(0)} (\vec r)= (0, xB_0,0)$, so that, in the presence
of an electrostatic potential, the Hamiltonian can be written as
\begin{equation}
H = H^{(0)} + H^{\rm m} +H^{\rm em} \, , \label{htot}
\end{equation}
where
\begin{equation}
H^{(0)}=\frac{1}{2m} \left[ p_x^2 +\left(p_y +\frac{e}{c}xB_0\right)^2\right]
\label{h_0}
\end{equation}
is the Hamiltonian of the homogeneous system with the usual Landau
eigenenergies $E_n^{(0)}= \hbar \omega_c (n+1/2)$ and 
eigenstates $|n, k_y \rangle $ centered around $x_0=-l^2
k_y$. The other two terms are Fourier sums as in
Eq.(\ref{fourier}). For the term linear in the magnetic modulation we have
\begin{equation}
 H^{\rm m}_{\vec q} = \frac{i}{q^2}  \left[ q_y p_x-q_x(p_y +
   m \omega_c x) \right] \, \omega_{\vec q} \, , \label{hmag}
\end{equation}
where we have introduced $ \omega_{\vec q} = eB_{\vec q}/mc$ in analogy to
the cyclotron frequency $ \omega_c = eB_0/mc$ of the homogeneous system. The 
 quadratic term in the magnetic modulation can be rewritten as a single
 Fourier expansion, and then
has the same form as the elctrostatic modulation, to which it adds as
\begin{equation}
H^{\rm em}_{\vec q} = V_{\vec q} + M_{\vec q} \, , \label{he-m}
\end{equation}
with
\begin{equation}
M_{\vec q}= -\frac{m}{2} \sum _{\vec k} \frac{ (\vec q - \vec k)\cdot
  \vec k }{(\vec q - \vec k)^2 \,  k^2} \, \omega_{\vec q - \vec k} \,
\omega_{\vec   k} \, ,  \label{mquad}
\end{equation}
where the sum is over all ${\vec   k} =(m_x K, m_y K,0)$ with integers $m_x$
and $m_y$.
Note that $ \omega_{\vec q =0}=0$  whereas  $M_{\vec q
  =0}$ yields a constant energy shift which we neglect. The matrix
elements of $H^{\rm em}$
in the Landau basis are immediately given by the
known\cite{Pfannkuche92:12606} matrix elements $\langle n', k'_y | \exp (i
\vec q \cdot \vec r) |n,k_y \rangle $. Those of $ H^{\rm m}$ can also be reduced
to these  matrix elements after the action of the momentum and position
operators in Eq.~(\ref{hmag}) on the Landau functions has been
evaluated using the  recurrance relations of
the Hermite polynomials.
%\begin{equation}
%\langle n', k'_y | H^{\rm m} |n,k_y \rangle = -\frac{i}{\sqrt{2}} \sum_{\vec q}
% \frac{\hbar  \omega_{\vec q}}{l q^2} \left[ \sqrt{n+1} (q_x-iq_y) \langle n',
% k'_y | e^{i \vec q \cdot \vec r}  |n+1,k_y \rangle + \sqrt{n} (q_x +i q_y) 
%\langle n', k'_y | e^{i \vec q \cdot \vec r}  |n-1,k_y \rangle \right] \, ,
%  \label{matrixel}
%\end{equation}
It is easily seen that the Hamiltonian of Eq.(\ref{htot}) has the same
symmetry under lattice translations and gauge transformations as the
Hamiltonian in the case with purely electric modulation. As a consequence, in
order to calculate its eigenstates and -energies, we may take the same steps
as in the treatment of the purely electric Hofstadter
problem\cite{Pfannkuche92:12606}. First, we assume both the electric and the
magnetic modulation to be weak enough, so that the coupling of different
Landau levels may be neglected. Then, within one Landau level, we make for the
eigenstates the ansatz  
\begin{equation}
||n,\alpha \rangle \rangle = \sum_{\lambda =-\infty}^{\infty}
c_{\lambda}(n,\alpha) |n,k_y + \lambda K\rangle \, \label{ansatz}
\end{equation}
with the restriction $|k_y|<K/2$. The coefficients $c_{\lambda}(n,\alpha)$
must be determined from an eigenvalue equation
\begin{equation}
\sum_{\lambda =-\infty}^{\infty} \left\{ H_{\lambda',\lambda} -E_{n,\alpha}
  \delta_{\lambda',\lambda} \right\} c_{\lambda}(n,\alpha) =0 \, .
  \label{eigengl}
\end{equation}
The matrix elements of $H$ are of the form $H^{(0)}_{\lambda',\lambda}=
\delta_{\lambda',\lambda} E^{(0)}_n$, and, for the
modulation-induced terms,
\begin{equation}
H^{\mu}_{\lambda',\lambda}=\sum_{\vec q} \delta_{\lambda',\lambda+
  \frac{q_y}{K}} \, 
{\cal S}^{\mu}_n ({\vec q})  \, e^{-il^2 q_x[k_y +\lambda K
  +\frac{q_y}{2}]} \, \label{hmll}
\end{equation}
with
\begin{equation}
{\cal S}^{\rm m}_n (\vec q )=
{\cal G}_n (Q) \, \hbar \omega_{\vec q} \, , \label{smag}
\end{equation}
\begin{equation}
{\cal S}^{\rm em}_n (\vec q )=
{\cal L}_n (Q) \, (V_{\vec q}+ M_{\vec q} )  \, . \label{se-m}
\end{equation}
Here we use the notation\cite{Pfannkuche92:12606} $Q=l^2 q^2 /2$, ${\cal L}_n
(Q)= \exp (-Q/2) \,  L_n(Q)$,  and ${\cal G}_n (Q) =\exp (-Q/2)
\,[L^{(1)}_{n-1} (Q) +L_n (Q)/2]$, where $L^{(\nu)}_{n} (Q)$ is an
associated Laguerre polynomial [$L_n = L_n^{(0)}$,
$L^{(\nu)}_{-1}=0$]. Equation (\ref{eigengl}) is a generalization of Harper's
equation, which contains arbitrary Fourier components of the periodic magnetic
and electrostatic fields. 

We now retain only the simplest Fourier
contributions necessary to describe typical experimental situations. First, we
assume that the (stress induced) electrostatic potential has inversion
symmetry with respect to the center of a magnetized post, i.e. has a pure
cosine expansion with real coefficients $ V_{\vec q}$. Second, we consider
only a simply harmonic modulation of the magnetic field in both directions,
$\omega_{\vec g _x }=\omega _x $ for $\vec g _x = (K,0)$ and
$\omega_{\vec g _y }=\omega _y $ for $\vec g _y = (0,K)$,
and, of course $\omega_{ - \vec q}= \omega^*_{\vec q}$, and
$\omega_{\vec q }=0$ for all other values of ${\vec q }$. This implies finite
contributions for $M_{ \vec q}$ with $\vec q =  \pm 2 \vec g _x$ and $ \pm 2
\vec g _y$, and 
$M_{\vec q}=0$ for all other values of $\vec q$. Finally, we retain in
the electrostatic potential energy only contributions with the same
wavevectors, i.e. the ground and the second harmonics
$V_{\pm \vec g _{\mu}}$ and  $V_{\pm 2 \vec g _{\mu}}$ for $\mu = x$   or $y$,
and we assume that these coefficients are real. Then,
with $E_{n,\alpha}=\hbar 
\omega_c (n+1/2) +\epsilon_{n,\alpha}$,  the generalization
(\ref{eigengl}) of Harper's equation reads
\begin{eqnarray}     \nonumber
\left[  2 \, {\rm Re} \left( \tilde{ {\cal S}}_n (\vec g _x) \, e^{-il^2 K [k_y +\lambda K]}
  +{\cal     S}^{\rm em}_{n}(2 \vec g _x) \, e^{-il^2 2K   [k_y 
  +\lambda K] } \right) - \epsilon_{n, \alpha} \right] \, c_{\lambda}\\
 +  \tilde{{\cal S}}_n (\vec g _y)  c_{\lambda +1}+\tilde{{\cal S}}_n (\vec g _y)^*  c_{\lambda -1}
+{\cal S}^{\rm em}_{n}(2 \vec g _y)   c_{\lambda +2}
+{\cal S}^{\rm em} _{n}(2 \vec g _y)^*  c_{\lambda -2} =0 \, , \label{harper}
\end{eqnarray}
where 
\begin{equation}
\tilde{{\cal S}}_{n}(\vec g _{\mu})=V_{\vec g _{\mu}} {\cal L}_n(Q) + \hbar
\omega_{\mu} \, {\cal G}_n(Q)\, , \label{stilde} \\ 
% {\cal S}^{(2)}_{\mu }&=&[V^{(2)}_{\mu } -(m/2)( \omega_\mu /K)^2 \,] \,
% {\cal L}_n(4Q) 
% \, , \label{smu2}
\end{equation}
for $\mu = x$ or $y$, and $Q=l^2 K^2 /2$. 

Let us first ignore the second-harmonic terms ${\cal S}^{\rm em}_n(2 \vec g
_{\mu})$. Then 
Eq.~(\ref{harper}) takes the usual form of Harper's
equation\cite{Pfannkuche92:12606}, however with complex coefficients $
\tilde{{\cal   S}}_{n}(\vec g _{\mu}) = |{\cal S}_{\mu}| \exp (i \varphi
_{\mu})$.  If one introduces $c_{\lambda} = \exp (-i \lambda \varphi_y) 
\tilde{c}_{\lambda}$,  the  $\tilde{c}_{\lambda}$ satisfy Harper's equation
with the real coefficients $ |{\cal S}_y |$ in the off-diagonal and 
$2 |{\cal S}_x | \cos (l^2 K [k_y +\lambda K] -  \varphi _x)$ in the diagonal
terms. 

Thus, for a modulation in $x$ direction only, i.e. for  $ {\cal S}_y
=0$, one obtains simple cosine bands of width $4 |{\cal S}_x |$. This
generalizes results obtained by Peeters and 
Vasilopoulos\cite{Peeters93:1466} for the special case of (i) real $\omega_x$,
i.e. a cosine expansion of the magnetic modulation field (``in-phase
modulation''\cite{Peeters93:1466}) as can be realized according to
Eq.~(\ref{dipolfeld}) by a magnetization in $z$ direction, and (ii) imaginary
$\omega_x$, i.e. a pure sinus expansion of $B_z (x)$ (``out-of-phase
modulation''\cite{Peeters93:1466}) as can be realized by a magnetization
parallel to the plane of the 2D electron system. 

For a square-symmetric modulation, $ \tilde{{\cal   S}}_{n}(\vec g _{y}) =
\tilde{{\cal   S}}_{n}(\vec g _{x})   $, the same 
bandwidth factor $ |{\cal S}_x |$ as for the 1D modulation can be factorized
out of Harper's equation. If the commensurability condition (\ref{fluxrat})
holds, a Bloch ansatz reduces the infinite set (\ref{harper}) to a closed set
of $p$ equations.
The energy eigenvalues can be written in the
form\cite{Pfannkuche92:12606} 
\begin{equation}
\epsilon_{n,\alpha}=|{\cal S}_x | \, \tilde{\epsilon} (k_x,k_y ;j) \, ,
  \label{energyfak}
\end{equation}
where  $\tilde{\epsilon} (k_x,k_y ;j)$ consists of $p$ subbands
($j$=1,...,$p$) defined in the first magnetic Brillouin zone $|k_x| \leq
\pi/qa, ~|k_y| \leq \pi /a$ and is the same for all Landau bands $n$, except a
possible shift in the repeated zone scheme due to the $n$-dependent phase
$\varphi_x$. The spectrum of the internal energy structure
$\tilde{\epsilon} (k_x,k_y ;j)$  plotted versus the inverse flux is thus just
Hofstadter's butterfly, Fig.~1(a). For a purely magnetic modulation this result
was already known\cite{Yoshioka87:448}. We see that, in the purely harmonic
approximation, the relative phase between magnetic and electric modulation
determines only the width of the Landau bands, which is proportional to the
$n$-dependent modulus of ${\cal S}_x$. The internal structure is not affected
by  this phase.

We now discusss the effect of the second-harmonic terms in Eq.~(\ref{harper}).
According to Eq.~(\ref{mquad}) the magnetic modulation always introduces such
terms, which may, however, be small for a weak, purely harmonic magnetic
modulation. But a second-harmonic electric modulation, which is 
known to be important for stress-induced modulations, will have the same
effect. A nontrivial phase difference between first and second order terms may
result, for instance, if the second order is dominated by an electric
modulation, which we assumed to have real Fourier coefficients, while the
first order is dominated by a magnetic modulation due to an in-plane surface
magnetization. 
Therefore, it is of principle interest to investigate how these terms
may change the  butterfly. If we write ${\cal S}^{\rm em}_{n}(2 \vec g
_{\mu})= |{\cal S}_{\mu 2}| 
\exp (2i\chi_{\mu} )$ and express the $c_{\lambda}$ in Eq.~(\ref{harper}) in
terms of the $\tilde{c}_{\lambda}$, the coefficient of $\tilde{c}_{\lambda
  +2}$ becomes $|{\cal S}_{y 2}|\, \exp (2i \Delta \chi_y)$ and the
corresponding diagonal term becomes $2 |{\cal S}_{x 2}|\, \cos 2(l^2 K [k_y +
\lambda K] -\varphi_x - \Delta \chi_x)$, with $\Delta \chi_{\mu} = \chi_{\mu}
-\varphi_{\mu}$. In Fig.~1 we have plotted some typical spectra for the case
of square symmetry, ${\cal S}_{y 2}={\cal S}_{x 2}$, where only a single
relevant phase difference $\Delta \chi$ remains. The normalization of the
spectra is the same as in Fig.~1(a), i.e. the amplitude  of the
second-harmonic terms is chosen as $|{\cal S}_{x 2}/{\cal S}_x| =
C \exp (-3 \pi \Phi_0 / 2 \Phi)$, and $C$ is taken constant, appropriate for
the lowest Landau level, $n=0$.  Figure~1(b) is for $C=0.5$ and $\Delta \chi
=0$. The exponential factor suppresses the second-harmonic term as the inverse
flux ratio $\Phi_0 / \Phi$ approaches unity, the spectrum reduces to that of
the unperturbed butterfly in Fig.~1(a). This factor also destroys the
reflection symmetry about the value  $\Phi_0 / \Phi=0.5$. If we replace
$\Delta \chi =0$ by  $\Delta \chi = \pi /2$, i.e. change the sign in all
second-harmonic terms, the spectrum is reflected about the line of zero
energy (measured with respect to $\hbar \omega_c /2$). The spectrum in
Fig.~1(c) is for $\Delta \chi = \pi /2$, but for a large magnitude of the
second-harmonic term, $C=2$, and the asymmetry about the energy zero is
enhanced. 
Note the strong crossing of subbands for $\Phi_0 / \Phi<0.5$, which destroys
the large gap in the lower left part of Fig.~1(a). 
 Figure~1(d) is for $C=2$ and $\Delta \chi = \pi
/4$. It is nearly, but not exactly, reflection symmetric about the zero energy
line, as we found at this phase difference also for other values of $C$. Thus
we see that the relative phase between the Fourier coefficients of the
fundamental modulation and its second harmonic does not only affect the
($n$-dependent) widths of the Landau bands, but also manifestly their internal
structures. 

Although the details of the energy spectrum, and as a consequence of the
density of states,  are changed by the second-harmonic
terms, and in general their symmetry is lowered,  the overall gap structure
remains similar. Thus, qualitative consequences of the gap structure, 
for example the suppression of the band conductivity in 2D superlattices as
compared with 1D modulations, should remain valid in superlattices defined by
mixed electrostatic and magnetic modulations even in the presence of second
harmonics. Other harmonics and the coupling of different Landau levels, which
have been investigated for purely electrostatic
modulations\cite{Pfannkuche92:12606,Kuehn93:13019} and have the tendency to
lower the symmetry of the Hofstadter-type spectra, are not expected to change
the picture qualitatively.

This work was supported by the NATO collaborative research grant No. 921204.
One of us (D. P.) gratefully acknowledges financial support by the Deutsche
Forschungsgemeinschaft.

%
%

%
%\newpage
%
\begin{figure}
\caption{Scaled energy spectra, $\epsilon_{n, \alpha} / {\cal S}_x$, versus
  inverse flux ratio $\Phi_0   / \Phi$ for a 
  mixed magnetic and electric modulation with 
  square symmetry, in the $n=0$ Landau level. The relative
  strength  of the second-harmonic terms is given by
$|{\cal S}_{x 2}/{\cal S}_x| =C \exp (-3 \pi \Phi_0 / 2 \Phi)$, the relative 
phase by $\Delta \chi = \frac{1}{2}\arg ( {\cal S}_{x 2} / {\cal S}_x ^2)
$. The parameter values are 
(a) $C=0$, $\Delta \chi =0$, 
(b) $C=0.5$, $\Delta \chi =0$, 
(c) $C=2.0$, $\Delta \chi = \pi /2$, 
(d) $C=2.0$, $\Delta \chi = \pi /4$.
}     \label{figsc1}
\end{figure}

\end{document}